\documentclass[twocolumn]{svjour3}          
\smartqed  
\usepackage{graphicx}
\journalname{.}
\begin{document}
\title{Design and implementation of a tunable composite photonic crystal cavity on an optical nanofiber}
\author{Ramachandrarao Yalla$^{1}$   \and Kohzo Hakuta $^{2}$}
\institute{\at
$^{1}$School of Physics\\
University of Hyderabad\\
Hyderabad, Telangana, India-500046\\
              \email{ramu@uohyd.ac.in}           
           \and
           \at
$^{2}$Center for Photonic Innovations\\
University of Electro-Communications\\ 
Chofu, Tokyo, Japan-182-8585\\
\email{k.hakuta@cpi.uec.ac.jp}}
\date{}
\maketitle
\begin{abstract}
We report a novel approach to the design and implementation of a tunable cavity on an optical nanofiber (ONF). The key point is to create a composite photonic crystal cavity (CPCC), by combining an ONF and a chirped period defect mode grating. Using numerical simulations we design the CPCC with low scattering loss while tuning the cavity resonance wavelength of $\pm$$10$ nm around the designed wavelength. We experimentally demonstrate the tunability of the CPCC, showing good agreement with the simulation results. Our results lay the foundation for a versatile platform for ONF cavity-quantum-electrodynamics with narrow bandwidth quantum emitters.
\keywords{Nanophotonics and photonic crystals\and Photonic crystal waveguides}
\end{abstract}

\section{Introduction}
Nano-waveguides offer a versatile and growing platform for nano-photonics with various applications, typically in quantum optics \cite{Fam05,Kali08,ArnoTrap,KimTrap,Yalla12,Yu}, quantum photonics \cite{Review18}, and sensing \cite{SNC,LT}. Cavity creation on nano-waveguides is a crucial requirement for enhancing the light-matter interaction strength. To date, various approaches have been developed by directly fabricating nanostructures on the nano-waveguide itself \cite{Kali11,Thompson,Hausmann,Goban}. From the viewpoint of fiber networks, tapered optical fibers with sub-wavelength in diameter termed as optical nanofibers (ONFs) are particularly promising due to their ability of automatic coupling to single mode fibers \cite{Fam05,Kali08,ArnoTrap,KimTrap,Yalla12}.

\begin{figure*}[h]
\centering
 \includegraphics[width=0.75\textwidth]{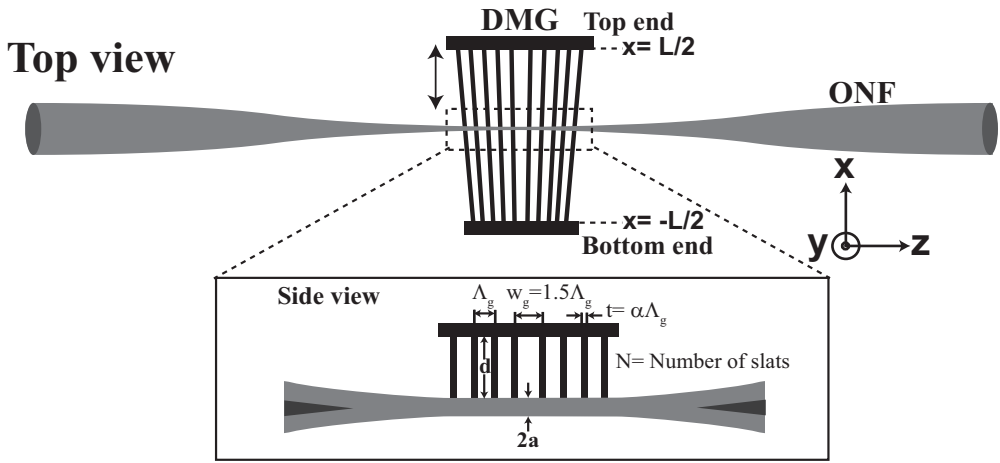}
\caption{{\label{Conceptual diagram}} A conceptual top view of a tunable composite photonic crystal cavity (CPCC). The tunable CPCC is formed by an optical nanofiber (ONF) and a chirped period defect mode grating (DMG). The bottom end and top end correspond to $x$= -$250$ $\mu$m (-$L$/$2$) and $x$= $250$ $\mu$m ($L$/$2$), respectively. The inset shows a conceptual side view of the tunable CPCC with defined parameters.} 
\end{figure*}

Cavity formation on the ONF has been demonstrated via two methods: one is the direct fabrication of photonic crystal cavities on the ONF itself using focused ion beam milling technique and femtosecond laser ablation \cite{Kali11,KaliPhC1,KaliPhC2,Schell15,Jam17,SNC1} and the other is a composite photonic crystal cavity (CPCC) method, which does not directly fabricate on the ONF. The CPPC was formed  by mounting the ONF onto a nanofabricated grating with a central defect \cite{Mark13,Yalla14,Jam15}. In order to extend such ONF cavities for cavity quantum electrodynamics (QED), one crucial requirement is the ability to tune the cavity resonance wavelength ($\lambda_{res}$) precisely to match with the narrow spectral emission line of a quantum emitter. Regarding the direct fabrication methods, tuning the $\lambda_{res}$-value up to $\pm$$10$ nm around the designed wavelength has been experimentally demonstrated by mechanical stretching of the ONF \cite{KaliPhC2,Schell15}. Regarding the composite method, although the tuning of the $\lambda_{res}$-value has not been experimentally demonstrated, two possible ways have been proposed. One is to precisely control the ONF diameter so that the effective refractive index of the ONF guided mode changes. The other is by changing the relative angle ($\theta$) between the ONF and the defect mode grating (DMG) \cite{Yalla14}. The precise control of the ONF diameter can be achieved via tensile and smooth tapered properties of the ONF. However, in order to tune the $\lambda_{res}$-value up to $\pm$$10$ nm, the expected ONF diameter variation should be $\pm$$6$\% \cite{Yalla14,Jam15}, leading to exceed the limit of tensile property of the ONF which may result in ONF break during the experiments. The smooth tapered ONF is not preferable, in the context of working with a solid state quantum emitter as it is deposited at a specific location on the surface of the ONF. The precise control of the $\theta$-value can be achieved via the rotation of either the ONF or the DMG. The required $\theta$-value should be $\pm$$10^\mathrm{o}$ to tune the $\lambda_{res}$-value up to $\pm$$10$ nm, leading to an increase of the scattering loss and resulting degradation in the performance over the tuning range. 

In this paper, we investigate a systematic design and implementation of tunable CPCC on the ONF, by combining the ONF and a chirped period DMG. As conceptually displayed in Fig.~\ref{Conceptual diagram} (top view), the essential point of the idea is to fabricate the DMG, the grating period of which is varied linearly from the bottom end to the top end. By changing the mounting position of the DMG onto the ONF from a minimum period position (bottom end) to a maximum period position (top end), the created CPCC $\lambda_{res}$-value can be tuned from one wavelength corresponding to the minimum grating period to another wavelength corresponding to the maximum grating period. 
\section{Tunable CPCC parameters design procedure}
We describe the design of a chirped period DMG and created CPCC. Here we restrict the discussion to a symmetric cavity structure. The directions $x$, $y$, and $z$ are defined as shown in Fig.~\ref{Conceptual diagram}. The essential parameters are schematically illustrated in the inset of Fig.~\ref{Conceptual diagram} (side view). Design parameters for the present CPCC are ONF diameter ($2a$), grating period ($\Lambda_g$), defect-width ($w_g$)= 1.5$\Lambda_g$, duty cycle ($\alpha$), slat width ($t$)= $\alpha\Lambda_g$, and number of slats ($N$). We assume a rectangular slat shape with a slat depth ($d$) of 2 $\mu$m. In the present design, we set the DMG length ($L$) to be $500$ $\mu$m. The DMG center is $x$= $0$. The bottom end and the top end correspond to $x$= -$250$ $\mu$m (-$L$/$2$) and $x$= $250$ $\mu$m ($L$/$2$), respectively. As conceptually depicted in Fig.~\ref{Conceptual diagram}, the position ($x$) where the DMG touches the ONF is defined as the mounting position ($x_m$). Thus, the $x_m$= -$250$ $\mu$m,  $x_m$= $0$, and $x_m$= $250$ $\mu$m correspond to the bottom end, center, and the top end, respectively. We design the present CPCC tunability of $\pm$$10$ nm around a specific wavelength at the center wavelength of $640$ nm. The $\Lambda_g$-value must varied linearly by $\pm$$5$ nm over the DMG length from the bottom end to the top end. 

\begin{figure}[h]
\centering
\includegraphics[width=8.5cm]{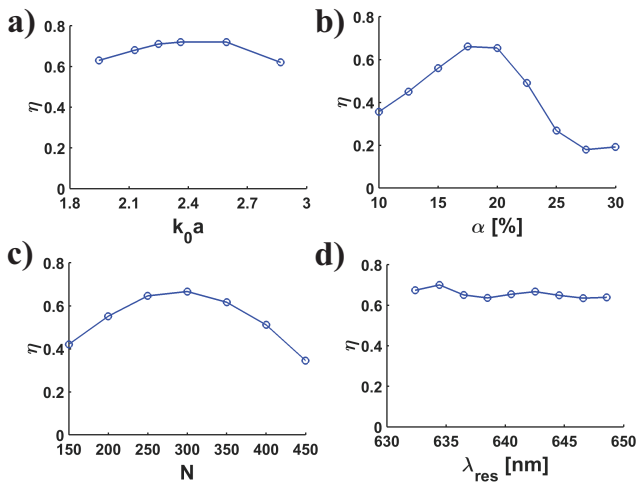}
\caption{{\label{Device design}}(a) Simulation predicted values for the channeling efficiency ($\eta$) as a function of fiber size parameter ($k_{0}$a). (b) $\eta$-value dependence on the duty cycle ($\alpha$) of the DMG. (c) $\eta$-value dependence on the number of slats ($N$) of the DMG. (d) $\eta$-value versus cavity resonance wavelength ($\lambda_{res}$).}
\end{figure}

Using the finite difference time domain (FDTD) method, we find the optimum parameters for the CPCC by simulating the channeling efficiency ($\eta$) into the ONF guided modes \cite{Yalla14}. We set a $y$-polarized dipole source on the surface of the ONF as it is expected to have maximum $\eta$-value. The parameter optimization procedure is as follows: according to Bragg resonance condition, $\lambda_{res}$= 2$n_{eff}\Lambda_g$, where $n_{eff}$ is the effective refractive index of the fundamental mode of the nanofiber and $\Lambda_g$ is the grating period. The effective refractive index depends on the nanofiber diameter ($2a$), the refractive indices of the core ($\sim1.45$) and clad ($\sim1$), and the wavelength of the light. We assume the slat width ($t$= $\alpha\Lambda_g$) to be around $50$ nm due to fabrication reliability as discussed in the Refs.$19$ and $20$. From the above, we obtain the relation $n_{eff}/\alpha= \lambda_{res}/100 $. The obtained $n_{eff}/\alpha$-value to be around $8$ assuming $\lambda_{res}$-value to be $800$ nm as discussed in Ref.$20$. The expected $\alpha$-value range would be $12.5$-$18.0\%$ assuming the $n_{eff}$-value range to be $1$-$1.45$. We simulated $\eta$-values for various  $n_{eff}$-values i.e. various $2a$-values (fiber size parameters, $k_{0}$a) to obtain the maximum $\eta$-value, while keeping the $\alpha$-value at $15$\% (average value of the range). It should be mentioned that $\Lambda_g$-value is chosen to produce $\lambda_{res}$-value at the designed wavelength and also $N$-value is swept to optimize the $\eta$-value. The summary of such results for the $\eta$-values versus $k_{0}$a-values are plotted in Fig.~\ref{Device design}(a). One can readily see that $\eta$-value is almost uniform over the $k_{0}$a-value from $2.2$ to $2.6$. We set the $k_{0}$a-value to be $2.5$ to minimize the scattering loss due to the fabricated slat shape deviation from a rectangular shape as discussed in Ref.$20$. Assuming the design wavelength to be around $640$ nm, the corresponding $2a$-value is $510$ nm. 

For the current design, the obtained $n_{eff}/\alpha$-value is around $6.4$ assuming the $\lambda_{res}$-value to be $640$ nm.  The expected $\alpha$-value range would be $15.5$-$22.5\%$. By monitoring the $\eta$-value at $x_m$= $0$, the $\alpha$-value is swept by $10$-$30$\%, while keeping $2a$-value fixed at $510$ nm i.e. $n_{eff}$-value is fixed. The simulated results are plotted in Fig.~\ref{Device design}(b), the optimum $\alpha$-value is found to be around $20$\%. The $N$-value is swept from $150$-$450$, while keeping the $\alpha$-value at the optimum. The simulated results are plotted in Fig.~\ref{Device design}(c), the optimum $N$-value is found to be around $300$. The $\Lambda_g$-value is swept to produce the designed $\lambda_{res}$-value ($640$ nm) and it is found to be $252$ nm. Thus, the minimum and maximum $\Lambda_g$-values are $247$ nm (bottom end) and $257$ nm (top end), respectively. Note that the angle difference between the center slat and the furthest slat is quite small $0.17^\mathrm{o}$, leading to negligible change in the scattering loss from parallel-slat DMGs. We also examined the $\eta$-value at various mounting positions ($x_m$) i.e. various $\lambda_{res}$-values. The simulated $\eta$-values versus $\lambda_{res}$-values are plotted in Fig.~\ref{Device design}(d). One can readily see that the $\eta$-value is uniform over the tuning range, suggesting stable performance of the present CPCC thanks to the small slat angle. Thus, we set the tunable CPCC parameters as follows: $2a$= $510$ nm, $\Lambda_g$= $252$ ($\pm5$) nm, $w_g$= $378$ ($\pm7.5$) nm, $\alpha$= $20$\%, $t$= $50.4$ ($\pm$1.0) nm, and $N$= $300$. 

\section{Experimental Procedure}

\begin{figure}[h]
\centering
 \includegraphics[width=0.49\textwidth]{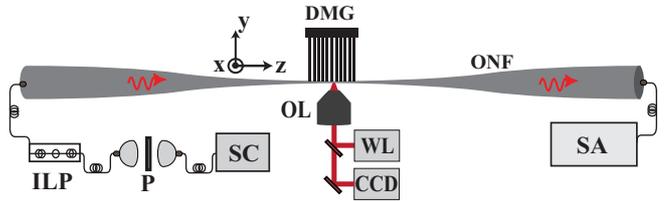}
\caption{{\label{Experimental Schematic}} The experimental setup for measuring the cavity transmission spectrum. SC, P, ILP, OL, WL, CCD, and SA denote super-continuum source, polarizer, in-line polarizer, objective lens, white light source, charge-coupled device, and spectrum analyzer, respectively.} 
\end{figure}
Based on the optimized parameters discussed above, chirped period DMG patterns were fabricated on a silica substrate using electron beam lithography along with chemical etching \cite{Mark13,Yalla14}. The ONF was fabricated using a heat and pull technique \cite{Jam17,Jam15}. The diameter of the ONF was $515$$\pm$5 nm. The optical transmission through the ONF was $97$\%. The ONF was placed at the focus point of an inverted microscope using a high precision {\it{xz}}-stage. By observing the DMG pattern using a CCD camera, the ONF was positioned perpendicular to the slats using a rotational stage. Details can be found in Ref.19 and 20. The position of the $x_m$= $0$ was determined by finding the top and bottom edge of the DMG pattern.  

The experimental setup for the optical characterization of the present CPCC is shown in Fig.~\ref{Experimental Schematic}. We inject a filtered supercontinuum light source (SC) spanning from $600$ nm to $700$ nm. The light is directed through a polarizer (P) to ensure linear polarization. The injected light polarization angle is controlled via a fiber in-line polarizer (ILP). The resultant cavity transmission spectrum is measured using a spectrum analyzer (SA) with a resolution of $0.05$ nm.
\section{Simulation and Experimental Results}
\begin{figure*}
\centering
\includegraphics[width=10.8cm]{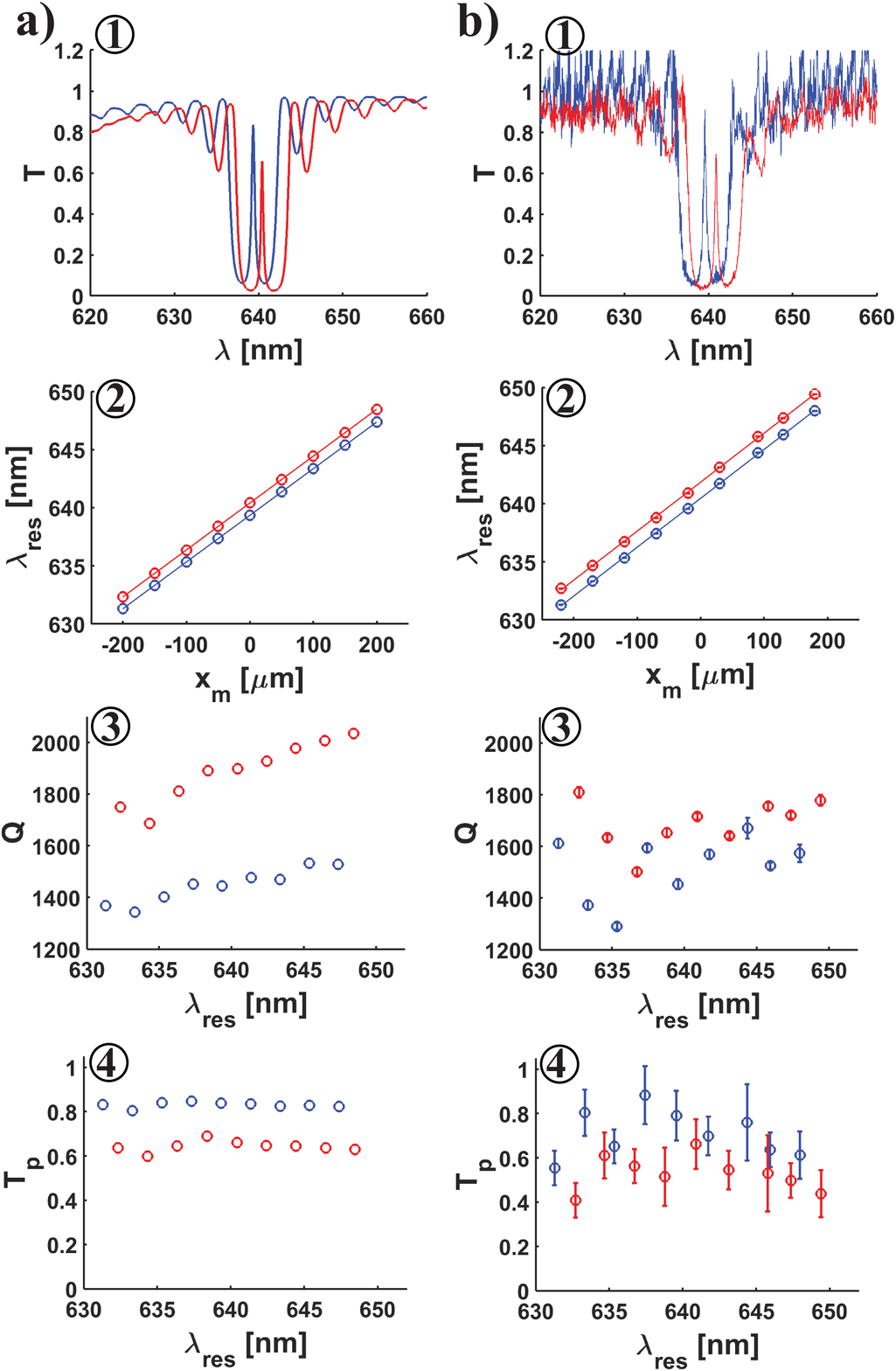}
\caption{{\label{Results}}Simulated and corresponding measured results for a tunable composite photonic crystal cavity (CPCC): (a) Simulated results are labeled from $1$ to $4$. (a1) Cavity transmission spectra for the $x$ (blue trace) and $y$ (red trace)-modes at the mounting position ($x_m$)= $0$, respectively. (a2) Cavity resonance wavelength ($\lambda_{res}$) as a function of $x_m$-values. Blue (red) circles correspond to the $x$ ($y$)-mode. Blue (red) solid line is a linear fit to the $x$ ($y$)-mode $\lambda_{res}$-values. (a3) and (a4) are the obtained quality factor ($Q$) and peak transmission ($T_p$) versus $\lambda_{res}$-values, respectively. In both the plots, blue (red) circles correspond to the $x$ ($y$)-mode. (b) The measured results corresponding to the simulation predicted results are labeled from $1$ to $4$. (b1) Cavity transmission spectra for the $x$ (blue trace) and $y$ (red trace)-modes at $x_m$= -$20$ $\mu$m, respectively. (b2) The $\lambda_{res}$-values as a function of $x_m$-values. Blue (red) circles correspond to the $x$ ($y$)-mode. Blue (red) solid line is a linear fit to the $x$ ($y$)-mode $\lambda_{res}$-values. (b3) and (b4) show the obtained $Q$-values and $T_p$-values versus $\lambda_{res}$-values, respectively. In both the plots, blue (red) circles correspond to the $x$($y$)-mode.}
\end{figure*}

The simulated and measured tunable CPCC characteristics are shown in Figs.~\ref{Results}(a) and (b), respectively. We simulated the cavity transmission spectra at various mounting positions ($x_m$) with a step size of  $\pm$$50$ $\mu$m from the $x_m$= 0. In Fig.~\ref{Results}(a)-$1$, we show a typical simulated cavity transmission spectra for both $x$- and $y$-polarizations at $x_m$= $0$. Blue (red) trace corresponds to the $x$ ($y$)-polarization. For both the traces, one can readily see a strong photonic stop-band at $640$ nm along with a peak at the center. Due to the asymmetrical index modulation, the degeneracy of the $x$- and $y$-polarized fundamental modes of the ONF is lifted. The $\lambda_{res}$-value is found to be $639.36$ nm ($640.42$ nm) for the $x$ ($y$)-mode. The $x$- and $y$-modes resonance peaks are separated by $1.06$ nm. The obtained values of quality factor ($Q$) and peak transmission ($T_p$) are $1443$ ($1898$) and $0.84$ ($0.66$) for the $x$ ($y$)-mode, respectively. The $y$-mode has a higher $Q$ (lower $T_p$)-value than the $x$-mode due to the fact that $y$-mode experiences more stronger modulation, leading to a higher reflectivity (scattering loss). Note that any apparent performance degradation in simulation predicted values is not observed in comparison with the previously reported values \cite{Yalla14}, where the parallel (non-tilted) slats were used. 

We mount the DMG around the center of the pattern i.e. $x_m$= $0$ to produce the $\lambda_{res}$-value around the designed wavelength ($640$ nm). The measured cavity transmission spectra corresponding to the simulated spectra is shown Fig.~\ref{Results}(b)-$1$. The measured cavity transmission spectra correspond to the $x_m$-value of -$20$ $\mu$m. Blue (red) trace corresponds to the $x$ ($y$)-mode. In both the traces, we observed a strong optical stop-band at a wavelength of $640$ nm accompanied by a single peak at the center as predicted by the simulations. The peak corresponds to the $\lambda_{res}$-value of $639.57$ nm ($640.90$ nm) for the $x$ ($y$)-mode. The separation in $\lambda_{res}$-values for the $x$- and $y$-modes is $1.33$ nm. The obtained values of $Q$ and $T_p$ are $1453$$\pm$$20$ ($1715$$\pm$$15$) and $0.79$$\pm$$0.11$ ($0.66$$\pm$$0.06$) for the $x$ ($y$)-mode, respectively. The measured stop-band, clear polarization dependence, and cavity mode around the designed wavelength ($640$ nm) behavior reproduce the simulation results as shown in Fig.~\ref{Results}(a)-$1$. Note that the measured $\lambda_{res}$-value, $Q$-value, and $T_p$-value for the the $x$- and $y$-modes are in good quantitative agreement with the simulation predicted values.

The simulated tunability of the present CPCC i.e. the obtained $\lambda_{res}$-values as a function of $x_m$-values are plotted in Fig.~\ref{Results}(a)-$2$. Blue (red) circles correspond to the $x$ ($y$)-mode. The  $\lambda_{res}$-value increases linearly with the $x_m$-value. This is due to a linear variation of the $\Lambda_g$-value from the bottom end to the top end of the DMG. We fit a linear function to the simulation predicted values. The fitted result for the $x$ ($y$)-mode $\lambda_{res}$-values is shown by blue (red) solid line. For the $x$ ($y$)-mode, the obtained slope ($S$) and resonance wavelength ($\lambda_0$) at $x_m$= $0$  are $0.0401$ nm/$\mu$m ($0.0404$ nm/$\mu$m) and $639.34$ nm ($640.40$ nm), respectively.

We measured the cavity transmission spectra at various $x_m$-values, by translating the DMG position along the {\it{x}}-axis with a step size of $\pm$$50$ $\mu$m from the $x_m$= -$20$ $\mu$m. The experimentally measured tunability behavior corresponding to the simulation result is shown in Fig.~\ref{Results}(b)-$2$. Blue (red) circles correspond to the $x$ ($y$)-mode. One can readily see a linear dependence of the $\lambda_{res}$-values for the $x$ ($y$)-mode as predicted by the simulations. We obtained $S$-value and $\lambda_0$-value, by fitting a linear function to the measured data. Blue (red) solid line is the fitted result for the $x$ ($y$)-mode $\lambda_{res}$-values. For the $x$ ($y$)-mode, the obtained $S$-value and $\lambda_0$-value are $0.0416$ nm/$\mu$m ($0.0422$ nm/$\mu$m) and $640.42$ nm ($641.84$ nm), respectively. Note that the measured $\lambda_{res}$-values for the $x$ ($y$)-mode behavior reproduced the simulation predicted behavior as shown in Fig.~\ref{Results}(a)-$2$. We measured the tunability behavior for the nine DMG patterns using the same ONF. We found mean values of $S$ and $\lambda_0$ are $0.04190$$\pm$$0.00032$ nm/$\mu$m ($0.04210$$\pm$$0.00015$ nm/$\mu$m) and $640.57$ $\pm$$0.16$ nm ($642.09$$\pm$$0.21$ nm) for the $x$ ($y$)-mode, respectively.  

Next, we show the performance stability of the present CPCC over the tuning range. The obtained $Q$-values against $\lambda_{res}$-values are plotted in Fig.~\ref{Results}(a)-$3$. Blue (red) circles correspond to the $x$ ($y$)-mode. The $Q$-value increases with the $\lambda_{res}$-value from $1368$ ($1750$) to $1528$ ($2035$) for the $x$ ($y$)-mode. The obtained $Q$-values show fluctuation around the mean value of $1446$$\pm$$65$ ($1887$$\pm$$117$) for the $x$ ($y$)-mode. 

The measured $Q$-value versus $\lambda_{res}$-value corresponding to the simulation predicted values are shown in Fig.~\ref{Results}(b)-$3$. Blue (red) circles correspond to the $x$ ($y$)-mode. The vertical error bars are due to the fluctuations in the $\lambda_{res}$-value and the cavity mode width. The measured $Q$-values show fluctuation around the mean value of $1517$$\pm$$122$ ($1689$$\pm$$93$) for the $x$ ($y$)-mode. The measured $Q$-values for the $y$-mode are higher than the $x$-mode as predicted by the simulations. The measured behavior reproduced the simulation predicted results as shown in Fig.~\ref{Results}(a)-$3$. 

The simulated $T_p$-values versus the $\lambda_{res}$-values are plotted in Fig.~\ref{Results}(a)-$4$. Blue (red) circles correspond to the $x$ ($y$)-mode. The $T_p$-value is almost kept constant over the tuning range for the $x$ ($y$)-mode, suggesting stable performance of the present CPCC owing to the small slat angle. Note that the uniform variation of the slat width ($t$) and the defect-width ($w_g$) from the bottom end to the top end of the DMG, leading to uniform scattering loss over the tuning range. The obtained $T_p$-values show fluctuation around a mean value of $0.83$$\pm$$0.01$ ($0.64$$\pm$$0.02$) for the $x$ ($y$)-mode.

The measured $T_p$-values versus $\lambda_{res}$-values corresponding to the simulation predicted values are shown in Fig.~\ref{Results}(b)-$4$. Blue (red) circles correspond to the $x$ ($y$)-mode. The vertical error bars are mainly due to the fluctuations in the injected source used for the measurement. The measured $T_p$-values show fluctuation around the mean value of $0.71$$\pm$$0.10$ ($0.53$$\pm$$0.08$) for the $x$ ($y$)-mode. The measured $T_p$-values for the $y$-mode are lower than the $x$-mode as predicted by the simulations. 

\section{Discussion}
Regarding the slope of the cavity tuning, $S$, the deviation from the simulation predicted value is about $4.5$\% ($4.2$\%) for the $x$ ($y$)-mode. This may be due to the fabrication imperfections in the grating period ($\Lambda_g$) of the DMG. Regarding the $\lambda_0$-value, the discrepancy from the simulation predicted value is about $1.23$ nm ($1.69$ nm) for the $x$ ($y$)-mode. This may be due to the fabrication fluctuations in the nanofiber diameter ($2a$) and the $\Lambda_g$-value. We measured the $2a$-value using scanning electron microscope and confirmed it to be close to the simulation set value of $510$ nm. It should be mentioned that the measurement accuracy in $2a$-value is $\pm$$5$ nm. Using the measured $\lambda_{0}$-values and assuming the fluctuations in the $\Lambda_g$-value to be around $\pm$$0.5$ nm, we estimate the $2a$-value to be $514$$\pm$$4$ nm \cite{Yalla14,Jam15}. This implies that the nanofiber diameter ($2a$) is thicker than the simulation set value. 

Regarding the separation in $\lambda_{0}$-values for the $x$- and $y$-modes, the discrepancy from the simulation predicted value is about $0.4$ nm. This may be due to thicker nanofiber diameter, which was used for current experiments. Simulations suggest that the separation is dependent on the $2a$-value. By setting the $2a$-value to be $520$ nm, the simulation predicted value for the separation is $1.49$ nm. 

Although the experimental discrepancies exist, it should be mentioned that the obtained $S$-value and $\lambda_{0}$-value are in good agreement with the simulation predicted values within the experimental errors. The measured and simulation results clearly demonstrate that the tunability for the present CPCC is about $\pm10$ nm around the designed wavelength of $640$ nm. 

Regarding the measured $Q$-values, the deviation from the simulation predicted values are about $1$-$17$\% ($3$-$17$\%) for the $x$ ($y$)-mode. This may be due to the fabrication fluctuation in the slat width ($t$) and the defect-width ($w_g$) over the DMG length, leading to a non-uniform scattering loss. The measured $T_p$-values are deviated from the simulation predicted values by about $1$-$33$\% ($1$-$36$\%) for the $x$ ($y$)-mode. This may be due to the fabrication imperfections in the slat width ($t$) and the defect-width ($w_g$) over the DMG length, leading to a non-uniform scattering loss. Although the fluctuations in the measured results, the simulation and experimental results clearly demonstrate that the stable performance of the present CPCC over the tuning range. 

Using the measured $S$-value, we estimate the precision of the tuning is $7.6$ GHz ($10.4$ pm), considering the ONF movement resolution along the {\it{x}}-axis to be $250$ nm. The tuning precision can be improved by increasing the ONF movement resolution. Note that the current tuning precision is about $1/35$ factor of the measured cavity $y$-mode width ($0.374$ nm). We believe that such a value would be good enough to perform any cavity-QED experiments. On the other hand, for some applications wider tunability is necessary. Simulations suggest that, choosing a larger slat angle ($\theta_{s}$) of $0.85^\mathrm{o}$, we could achieve the tunability up to $\pm50$ nm around the designed wavelength. Note that the present results can be readily extended to various designed wavelengths. 

\section{Summary}
In this paper, we have demonstrated the design and implementation of tunable composite photonic crystal cavity on an optical nanofiber. The numerical and experimental results have clearly shown that the composite cavity method can be extended to a tunable cavity scheme without suffering from any additional scattering loss. Experimental results have reproduced the simulated results successfully. Although discussions were restricted to a symmetric cavity structure, this method can be extended to any asymmetric cavity structure, such as one-side cavity. The present method can readily be applied to ONF cavity-QED works with narrow bandwidth quantum emitters such as laser cooled atoms \cite{Mark17}, quantum dots at cryogenic temperatures \cite{Biadala09,Shafi}, and silicon vacancy centers in nano-diamonds \cite{Neu}, and may open new avenues and lay a versatile platform in the fields of quantum optics and nano-photonics.

\begin{acknowledgements}
This work was supported by the Japan Science and Technology Agency (JST) as one of the strategic innovation projects.
\end{acknowledgements}
\section*{Conflict of interest}
The authors declare that they have no conflict of interest.


\begin{thebibliography}{}
\bibitem{Fam05} 
F. L. Kien, S. Dutta Gupta, V. I. Balykin, and K. Hakuta, Phys. Rev. A \textbf{72}, 032509 (2005). 
\bibitem{Kali08} 
K. P. Nayak and K. Hakuta, New J. Phys. \textbf{10}, 053003 (2008).
\bibitem{ArnoTrap} E. Vetsch, D. Reitz, G. Sagu\'{e}, R. Schmidt, S. T. Dawkins, and A. Rauschenbeutel, Phys. Rev. Lett. \textbf{104}, 203603 (2010).
\bibitem{KimTrap} A. Goban, K. S. Choi, D. J. Alton, D. Ding, C. Lacroute, M. Pototschnig, T. Thiele, N. P. Stern, and H. J. Kimble, Phys. Rev. Lett. \textbf{109}, 033603 (2012).
\bibitem{Yalla12} R. R. Yalla, F. L. Kien, M. Morinaga, and K. Hakuta, Phys. Rev. Lett. {\bf{109}}, 063602 (2012).
\bibitem{Yu} S.-P. Yu {\it{et al.}}, Appl. Phys. Lett. {\bf{104}}, 111103 (2014).
\bibitem{Review18} K. P. Nayak, M. Sadgrove, R. R. Yalla, F. L. Kien, and K. Hakuta, J. Opt. {\bf{20}}, 073001 (2018).
\bibitem{SNC} M. J. Morrissey, K. Deasy, M. Frawley, R. Kumar, E. Prel, L. Russell, V. G. Truong, and S. N. Chormaic, Sensors \textbf{13}, 10449 (2013).
\bibitem{LT} J. Lou, Y. Wang, and L. Tong, Sensors  \textbf{14}, 5823 (2014).
\bibitem{Kali11} K. P. Nayak, F. L. Kien, Y. Kawai, K. Hakuta, K. Nakajima, H. T. Miyazaki, and Y. Sugimoto, Opt. Express \textbf{19}, 14040-14050 (2011).
\bibitem{Thompson} J. D. Thompson, T. G. Tiecke, N. P. de Leon, J. Feist, A. V. Akimov, M. Gullans, A. S. Zibrov, V. Vuletic, and M. D. Lukin, Science \textbf{340},  1202 (2013).
\bibitem{Hausmann} J. M. Hausmann, B. J. Shields, Q. Quan, Y. Chu, N. P. de Leon, R. Evans, M. J. Burek, A. S. Zibrov, M. Markham,D. J. Twitchen, H. Park, M. D. Lukin, and M. Lonc\v{a}r, Nano Lett. \textbf{13}, 5791 (2013). 
\bibitem{Goban} A. Goban, C.-L. Hung, S.-P. Yu, J. D. Hood, J. A. Muniz, J. H. Lee, M. J. Martin, A. C. McClung, K. S. Choi, D. E. Chang, O. Painter, and H. J. Kimble, Nat. Commun. \textbf{5}, 4808 (2014).
\bibitem{KaliPhC1} K. P. Nayak and K. Hakuta, Opt. Express \textbf{21}, 2480 (2013).
\bibitem{KaliPhC2} K. P. Nayak, P. Zhang, and K. Hakuta, Opt. Lett. \textbf{39}, 232 (2014).
\bibitem{Schell15} A. W. Schell, H. Takashima, S. Kamioka, Y. Oe, M. Fujiwara, O. Benson, and S. Takeuchi, Sci. Rep. \textbf{5}, 9619 (2015).
\bibitem{Jam17} J. Keloth, K. P. Nayak, and K. Hakuta, Opt. Lett. \textbf{42}, 1003 (2017).
\bibitem{SNC1} W. Li, J. Du, V. G. Truong, and S. Nic Chormaic, Appl. Phys. Lett.  \textbf{110}, 253102 (2017).
\bibitem{Mark13} M. Sadgrove, R. R. Yalla, K. P. Nayak, and K. Hakuta, Opt. Lett. \textbf{14}, 2542 (2013).
\bibitem{Yalla14}R. R. Yalla, M. Sadgrove, K. P. Nayak, and K. Hakuta, Phys. Rev. Lett. {\bf{113}}, 143601 (2014).
\bibitem{Jam15} J. Keloth, M. Sadgrove, R. R. Yalla, and K. Hakuta, Opt. Lett. \textbf{40}, 4123 (2015).
\bibitem{Mark17}M. Sadgrove and K. P. Nayak, New J. Phys. \textbf{19}, 063003 (2017). 
\bibitem{Biadala09} L. Biadala, Y. Louyer, Ph. Tamarat, and B. Lounis, Phys. Rev. Lett. {\bf{103}}, 037404 (2009).
\bibitem{Shafi}K. M. Shafi, W. Luo, R. R. Yalla, K. Iida, E. Tsutsumi, A. Miyanaga, and K. Hakuta, Sci. Rep. {\bf{8}}, 13494 (2018).
\bibitem{Neu} E. Neu, D. Steinmetz, J. Riedrich-Moller, S. Gsell, M. Fischer, M. Schreck, and C. Becher, New J. Phys {\bf{13}}, 025012 (2011).
\end{thebibliography}
\end{document}